\documentclass[preprint, twocolumn]{aastex7}

\usepackage{physics}
\usepackage{caption}
\usepackage{subcaption}

\newcommand\PLUTO{\texttt{PLUTO} }
\newcommand\Alfven{Alfv\'{e}n }
\newcommand\Alfvenic{Alfv\'{e}nic }

\newcommand{\Bin}{B_\mathrm{in} }
\newcommand{\Bbkg}{B_\mathrm{bkg} }
\newcommand{\dB}{\delta B }
\newcommand{\vA}{v_\mathrm{A} }
\newcommand{\vAn}{v_{\mathrm{A},0} }
\newcommand{\vfn}{v_{\mathrm{fast},0} }
\newcommand{\MA}{M_\mathrm{A} }
\newcommand{\cs}{c_\mathrm{s} }
\newcommand{\csn}{c_{\mathrm{s},0} }
\newcommand{\CA}{C_\mathrm{A} }

\newcommand{\plane}[2]{$#1$\nobreakdashes-$#2$~plane}

\begin{document}

\title{Simulations of a Conducting Sphere Moving through Magnetized Plasma: \Alfven Wings, Slow Magnetosonic Wings, and Drag Force}

\author[0000-0002-0088-2563]{Nicholas J. Corso}
\affiliation{Department of Astronomy, Center for Astrophysics and Planetary Science, Cornell University, Ithaca, NY 14853, USA}
\email[show]{njc86@cornell.edu}

\author[0000-0002-1934-6250]{Dong Lai}
\affiliation{Tsung-Dao Lee Institute, Shanghai Jiao-Tong University, Shanghai,  China}
\affiliation{Department of Astronomy, Center for Astrophysics and Planetary Science, Cornell University, Ithaca, NY 14853, USA}
\email[show]{dl57@cornell.edu}

\begin{abstract}

Plasma-mediated interaction between astrophysical objects can play an important
  role and produce electromagnetic radiation in various binary systems, ranging
  from planet-moon and star-planet systems to binary compact objects. We
  perform 3D magnetohydrodynamic numerical simulations to study an ideal
  magnetized plasma flowing past an unmagnetized conducting sphere. Such flow
  generates magnetic disturbances and produces a drag force on the sphere, and
  we explore the corresponding drag coefficient as a function of the flow speed
  relative to \Alfven speed and the $\beta$ parameter of the background plasma.
  We find that the drag is generally well-described by the \Alfven wing model,
  but we also show that slow magnetosonic waves provide a correction through
  their own wing-like features. These give rise to a nontrivial dependence of
  the drag coefficient on the plasma $\beta$, as well as enhanced drag as the flow
  speed approaches the \Alfven speed.

\end{abstract}

\section{Introduction} \label{sec:intro}

Viscous drag on a sphere moving through a fluid is a classic problem in
hydrodynamics, from its analytical conception in Stokes's law
\citep{Stokes1851}, through advances to higher Reynolds number using
experimental and computational techniques (see \citealt{Tiwari+20} for a
comprehensive review). The problem is best summarized through the relationship
between the drag coefficient and Reynolds number, which reveals changes in
scalings as the flow transitions from laminar to vortex formation and
shedding to turbulence. Despite the idealized setup of this problem, it has found a
wide array of applications in the physical sciences. What happens if the fluid
is replaced by a magnetized plasma and the sphere is conducting? Is there a
similar construction that we may use to describe this type of flow?

This problem is relevant for a number of  astrophysical scenarios where highly
magnetized objects interact with weakly or unmagnetized bodies. In particular,
magnetic interaction may play an important role and lead to electromagnetic
radiation in various types of binaries, including planet-moon systems
\citep{Goldreich&Lynden-Bell69, Neubauer90, Bagenal&Dols20}, star-planet
systems \citep{Li+98, Preusse+06, Laine&Lin11, Lai12, Saur18, Strugarek23,
Lee&Owen25}, and binary compact objects \citep{Hansen&Lyutikov01, Wu+02,
Dall'Osso+06, Lai12, Wang+16, Most&Philippov22, Most&Philippov23,
Ressler+24, Mahlmann&Beloborodov25}. In most cases, the strength of the
radiation and its detectability are highly uncertain. While there are
appreciable qualitative differences across the systems listed above, there are
unifying themes which are useful to consider generally. For instance, the
development of a magnetic flux tube connecting the objects and its subsequent
stressing and flaring are discussed in the context of planet-moon interactions
\citep{Goldreich&Lynden-Bell69}, star-planet interactions \citep{Lanza13} and
binary compact object interactions \citep{Lai12, Most&Philippov22}.

The concept of this connecting flux tube can be traced back to the construction
of \cite{Drell+65}, who considered the model of an ideal conductor (treated as
a rectangular bar) moving with velocity $\vb{v_0}$ in a lossless plasma with a
uniform magnetic field $\vb{B_0}$, to emulate a satellite traveling through
Earth's magnetic field. Their analysis is two-dimensional (i.e. the body is
infinite in extent in the direction perpendicular to $\vb{v_0}$ and
$\vb{B_0}$), and assumes that the conductor has the same internal magnetic
field as the external field. In the linear regime, when the magnetic
perturbation
$\left|\delta\vb{B}\right|\sim\left(\flatfrac{v_0}{\vA}\right)B_0$ (where
$\vA$ is the \Alfven speed) is much less than $B_0$, they concluded that the
magnetic disturbances of the plasma would propagate as \Alfven waves dragged by
the background flow to produce wakes that they named ``\Alfven wings.'' The
power radiated through the \Alfven wings drags the conductor. \cite{Neubauer80}
extended this picture into the regime of $v_0\lesssim\vA$
in applying the \Alfven wing picture to the
system of Jupiter and Io, and included resistive effects. The author claimed
that the \Alfven wings form an effective conductance which helps to close the
circuit around Io. In particular, under highly magnetized
conditions, the \Alfven wave along the wing can bounce off Jupiter and
return to Io before the moon has moved away, a closed current system can
connect the two bodies, reminiscent of the unipolar inductor model proposed by
\cite{Piddington67} and \cite{Goldreich&Lynden-Bell69}. In general
astrophysical binary systems, stellar winds may overwhelm the \Alfven wings,
preventing the latter from connecting the objects and instead directing them
out to infinity, a case referred to as a pure \Alfven wing \citep{Neubauer98}.
In recent years, magnetic interactions between stars (including stellar winds)
and exoplanets have been studied using numerical simulations (\textit{e.g.}
\citealt{Strugarek+14, Strugarek+15, DeColle+25}), but a general understanding
of how a conducting body interacts with an external magnetized plasma and how
the drag force and dissipation depend on the plasma properties, the velocity,
and the internal magnetic field of the body, is lacking.

In this paper, we begin a series study of the interaction between a moving
conductor and an ideal magnetized plasma. We employ three dimensional
magnetohydrodynamic (MHD) simulations to find steady-state solutions for the
flow and develop a framework for characterizing the drag coefficient as a
function of the body's velocity and internal magnetic field and the properties
of the magnetic plasma. This paper examines the zero internal field cases; future papers will consider more general magnetic configurations. This paper is organized as follows: in Section
\ref{sec:method}, we introduce the numerical methods for our simulations and
the problem setup; in Section \ref{sec:results}, we present the results of a
fiducial simulation to develop the qualitative picture of the interaction, and
we show the observed scaling with the canonical parameters of the problem;
we summarize these results in Section \ref{sec:discuss}.

\section{Problem Setup and Method} \label{sec:method}

\begin{figure}[ht]
  \begin{center}
    \includegraphics[width=\columnwidth, trim={300 0 300 0}, clip]{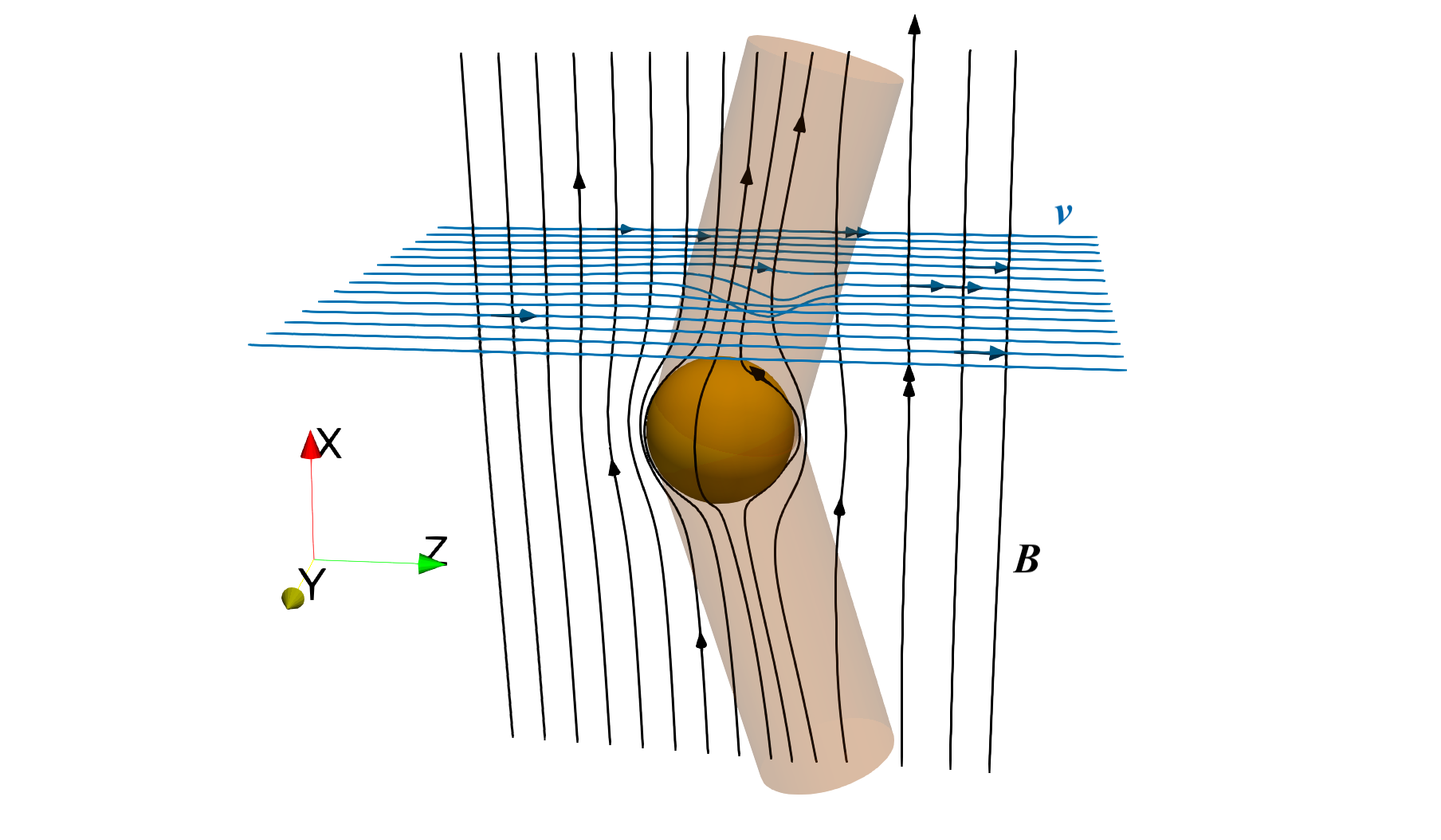}
    \caption{A sketch of the simulated system while it is in a steady state.
    The black lines are the magnetic field lines in the \plane{x}{z}, which
    asymptotically approach $\vb{B}=B_0\vu{x}$ for $r\gg R$, where $R$ is the radius of the conducting sphere. The
    blue lines are the velocity field lines in the $x=2R$ plane, which
    similarly approach an asymptotic value of $\vb{v}=v_0 \vu{z}$ for $r\gg R$. The orange sphere in the center represents the conductor,
    and the transparent orange cylinders emerging from it represent the
    theoretical locations of the \Alfven wing characteristics. \label{sketch}}
  \end{center}
\end{figure}

We consider a conducting sphere of radius $R$ moving with velocity
$\vb{v_0}=-v_0\vu{z}$ in a magnetized plasma. Far from the sphere ($r\gg R$),
the plasma density is $\rho_0$, and the magnetic field is $\vb{B}=B_0\vu{x}$.
In the rest frame of the conductor, the flow velocity is $\vb{v}=v_0\vu{z}$ at
$r\gg R$ (see Figure \ref{sketch}).

To simulate our system we use \PLUTO \citep{Mignone07}, a MHD code designed with a modular structure to handle a variety of
astrophysical contexts. In our case, we evolve the idealized MHD equations in
flux-conservative form,
\begin{align}
  \pdv{\rho}{t}+\div{(\rho \vb{v})} & =0, \\
  \pdv{\vb{m}}{t}+\div{\left(\vb{m} \vb{v}-\frac{\vb{\dB} \vb{B}}{4\pi}
  -\frac{\vb{\Bbkg} \vb{\dB}}{4\pi}\right)}+\grad{p_t} & =0, \label{momentum} \\
  \pdv{E_t}{t}+\div{\left[\left(E_t+p_t\right)
  \vb{v}-\frac{\left(\vb{v} \vdot \vb{\dB}\right) \vb{B}}{4\pi}\right]} & =0, \\
  \pdv{(\vb{\dB})}{t}-\curl{(\vb{v} \cp \vb{B})} & =0, \label{induction}
\end{align}
where $\rho$ is the plasma mass density, $\vb{v}$ is the velocity, $\vb{m}=\rho\vb{v}$
is the momentum density, $p$ is the gas pressure, and
\begin{equation}
  \label{B0} \vb{B}(\vb{r},t)=\vb{\Bbkg}(\vb{r})+\vb{\dB}(\vb{r},t).
\end{equation}
Here, $\vb{\Bbkg}(\vb{r})$ is a fixed background magnetic field (see below), and
$\vb{\dB}$ (which is not necessarily small) is an evolved difference from the
background field. The following are defined to represent the evolved terms for the
total pressure and energy density,
\begin{align} p_t & = p + \frac{\vb{\dB}^2}{8\pi} + \frac{\vb{\dB}\vdot\vb{\Bbkg}}{4\pi}, \\
              E_t & = \rho \varepsilon + \frac{1}{2}\left(\rho\vb{v}^2+\frac{\vb{\dB}^2}{4\pi}\right),
\end{align}
where $\varepsilon$ is the plasma internal energy per unit mass. We note that
\PLUTO itself adopts a convention in which the factor of
$\flatfrac{1}{\sqrt{4\pi}}$ is absorbed into the definition of the magnetic
field, but for this paper we will instead choose the CGS convention. These
equations are closed by an ideal equation of state, 
\begin{equation}
\rho \varepsilon = \frac{p}{\gamma-1},
\end{equation}
where we choose $\gamma=\flatfrac{5}{3}$, the adiabatic index of a monoatomic
ideal gas.

With this setup, the key dimensionless parameters of our system are the \Alfven
Mach number and the plasma $\beta$, defined by
\begin{equation}
\MA\equiv \frac{v_0}{\vAn},\quad\beta\equiv\frac{p_0}{B_0^2/4\pi}=\frac{\flatfrac{\csn^2}{\gamma}}{\vAn^2}, \label{params}
\end{equation}
where $\vAn=\flatfrac{B_0}{\sqrt{4\pi\rho_0}}$ and $\csn=\sqrt{\flatfrac{\gamma
p_0}{\rho_0}}$ are the \Alfven and sound speeds of the background flow
(associated with $\rho_0$, $p_0$, and $B_0$).

For all simulations, we use the second-order finite volume method that
evolves all quantities according to their cell-centered values. We employ a
third order weighted essentially non-oscillatory (WENO3) scheme for
reconstruction \citep{Yamaleev&Carpenter09} from cell centers to cell faces.
For most cases, we employ the linearized Roe Riemann solver for computing
fluxes at the cell faces \citep{Roe81}, but in some cases with high flow speeds
and low plasma $\beta$, we must rely on the more diffusive Harten-Lax-van Leer
Discontinuities (HLLD) Riemann solver \citep{Harten+83, Miyoshi&Kusano05}. We
use hyperbolic divergence cleaning to enforce the $\div{\vb{B}}$ condition
during evolution \citep{Dedner+02, Mignone&Tzeferacos10, Mignone+10}.

\subsection{Computational Domain and Boundary Conditions}

All simulations are run on a discretized grid using spherical coordinates
$(r,\theta,\phi)$, with a hard spherical conductor surface placed as the inner
boundary and the outer boundary $r_\mathrm{out}\gg R$ set to have a
zero-gradient outflow condition. Unless stated otherwise, the simulation domain
spans $r\in[R,10R]$ and the full unit sphere, using $N_r\times N_\theta\times
N_\phi=100\times64\times128$ grid points, where the radial points are spaced
logarithmically and the angular points uniformly. The polar axis coordinate
singularity is avoided due to the fact that cell-centered quantities are
evolved, but compression of cells near this boundary still prohibitively limits
the simulation speed due to the Courant–Friedrichs–Lewy (CFL) condition. To
avoid this problem,we follow \PLUTO and employ a ``ring-averaging'' technique,
in which cells close to the polar axis are combined to effectively increase
their size, increasing the azimuthal resolution by a factor of $2$ for every
step away from the polar axis until reaching the full resolution. For our
fiducial simulations, the cells adjacent to the polar axis effectively have an
azimuthal resolution of $N_\phi=16$. The boundary condition at the polar axis
is set such that cells receive information from across the pole, and the
azimuthal boundary condition is periodic.

Each simulation is initialized in the computational domain with a plasma of
mass density $\rho_0$ and pressure $p_0$ traveling with the velocity of a
potential flow,
\begin{equation}
\vb{v}=v_0\cos\theta\left[1-\left(\frac{R}{r}\right)^3\right]\vu{r} +
v_0\sin\theta\left[1+\frac{1}{2}\left(\frac{R}{r}\right)^3\right]\vu{\theta}.
\end{equation}
Note that at $r=R$ this already satisfies the hard boundary of the conductor
(\textit{i.e.} $v_r=0$), and far from the sphere the velocity approaches
$\vb{v}=v_0\vu{z}$. We assume that the conductor has no interior magnetic field
$\vb{\Bin}=0$. The magnetic field in the computational domain is
initialized such that it maintains continuity in $B_r$ across the conductor
boundary and asymptotically approaches $\vb{B_0}=B_0\vu{x}$ at $r\gg R$. The
combination of these two conditions specifies the fixed ``background'' field
$\vb{\Bbkg}$ introduced in Equation (\ref{B0}),
\begin{equation}
\vb{\Bbkg}=B_0\vu{x}-\left(\frac{R}{r}\right)^3\frac{B_0}{2}\left[3(\vu{r}\vdot\vu{x})\vu{r}-\vu{x}\right].
\end{equation}
The asymptotic behavior of $\vb{v}$ and $\vb{\Bbkg}$ can be seen in Figure
\ref{sketch}, which provides a schematic of the steady-state \Alfven wing
behavior. Our simulations start with the initial
$\delta\vb{B}=\vb{B}-\vb{\Bbkg}=0$.

Since the described initial conditions represent a background flow passing the
conductor, it is vital that this flow is preserved at large distances from it.
On its own, the zero-gradient outflow condition at $r_\mathrm{out}=10R$ could
result in divergent behavior since perturbations may provide feedback to the
background flow. As such, following the procedure of \cite{Laune+24}, we
introduce a damping zone in the region $8R<r<10R$. In this region, for a given
evolved variable in conservative form, $A$, which has a background value of
$A_0$, a damping term is introduced to the right hand side of the MHD
equations. This term, denoted $D_A$, takes the following form,
\begin{equation}
  D_A = -\zeta(x)\frac{A-A_0}{T_d},
\end{equation}
where the tapering function is given by
\begin{equation}
\zeta(x)= \begin{cases} 1-6 x^2+6 x^3 & 0 \leqslant x \leqslant \frac{1}{2} \\
                        2(1-x)^3 & \frac{1}{2}<x \leqslant 1, \end{cases}
\end{equation}
with
\begin{equation}
x=1-\frac{r-8R}{2R}.
\end{equation}
We set the damping timescale $T_d$ using the fastest characteristic speed
across the damping zone,
\begin{equation}
T_d\simeq\frac{2R}{\max\left(v_0, \vfn\right)},
\end{equation}
where $\vfn=\sqrt{\csn^2+\vAn^2}$ is the maximum value of the fast
magnetosonic wave speed.

\begin{figure*}[ht]
  \begin{center}
    \includegraphics[width=\textwidth, trim={0 100 0 100}, clip]{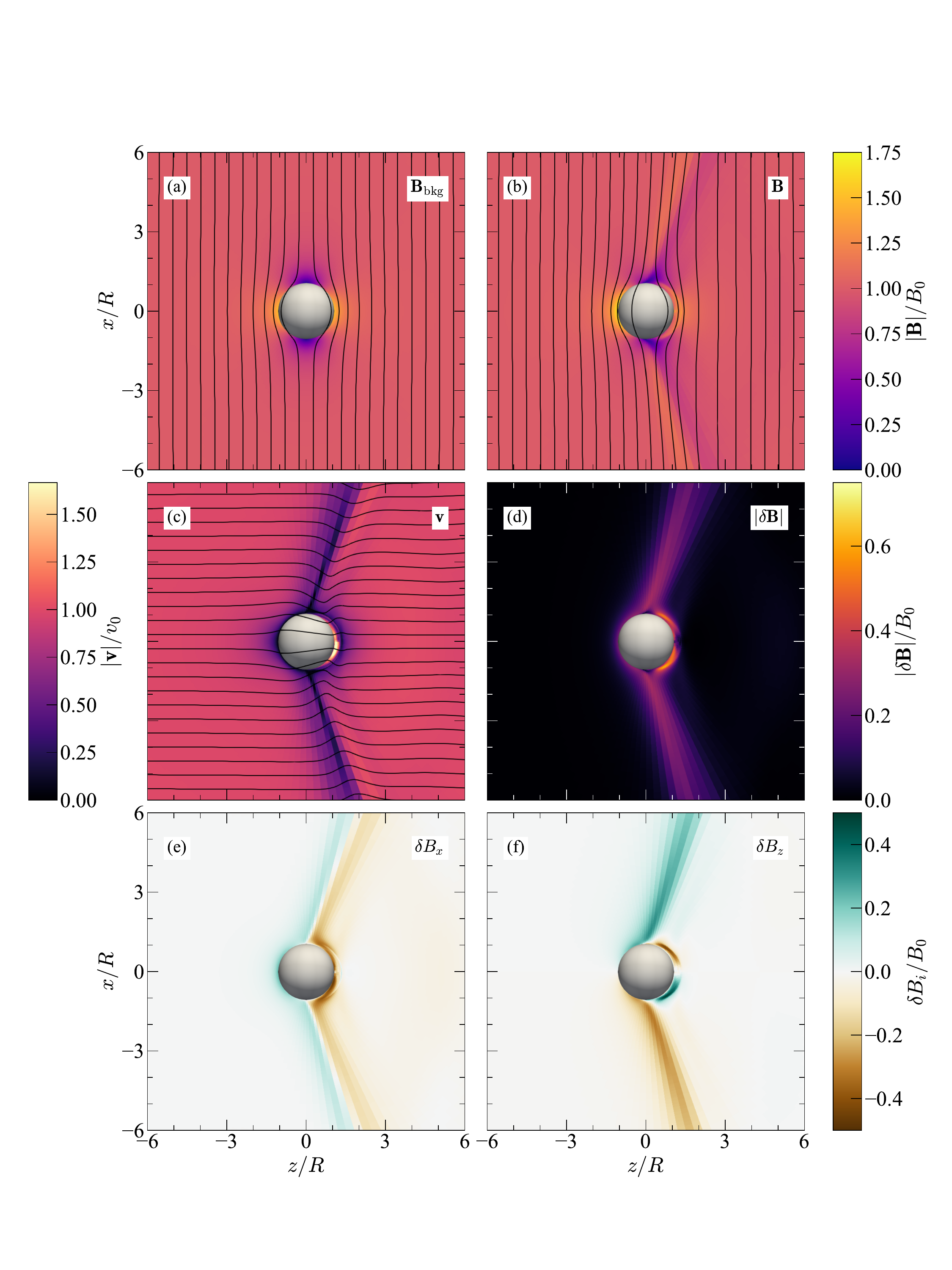}
    \caption{Diagnostic parameters for our fiducial run, with
    $\MA=0.3$, $\beta=2$, on the \plane{x}{z} after the
    simulation has achieved a steady state. The conductor is shown as a gray
    sphere in the center of the domain. Vector quantities are plotted with
    field lines placed in front of a background which displays magnitude. The
    field lines initialize at $y=0.15$ when $r=10R$ and may vary in $y$
    across the domain. From left-to-right, top-to-bottom, the quantities are
    (a) $\vb{\Bbkg}$, (b) $\vb{B}$, (c) $\vb{v}$, (d) $|\vb{\delta B}|$,
    (e) $\delta B_x$, and (f) $\delta B_z$.
    All evolved quantities show disturbances along the \Alfven wing
    characteristics. We note that the velocity field lines wrap around the
    wings since motion stops entirely along the characteristics. \label{fid_visual}}
  \end{center}
\end{figure*}
\begin{figure}[ht]
  \begin{center}
    \includegraphics[width=\columnwidth, trim={100 50 25 100}, clip]{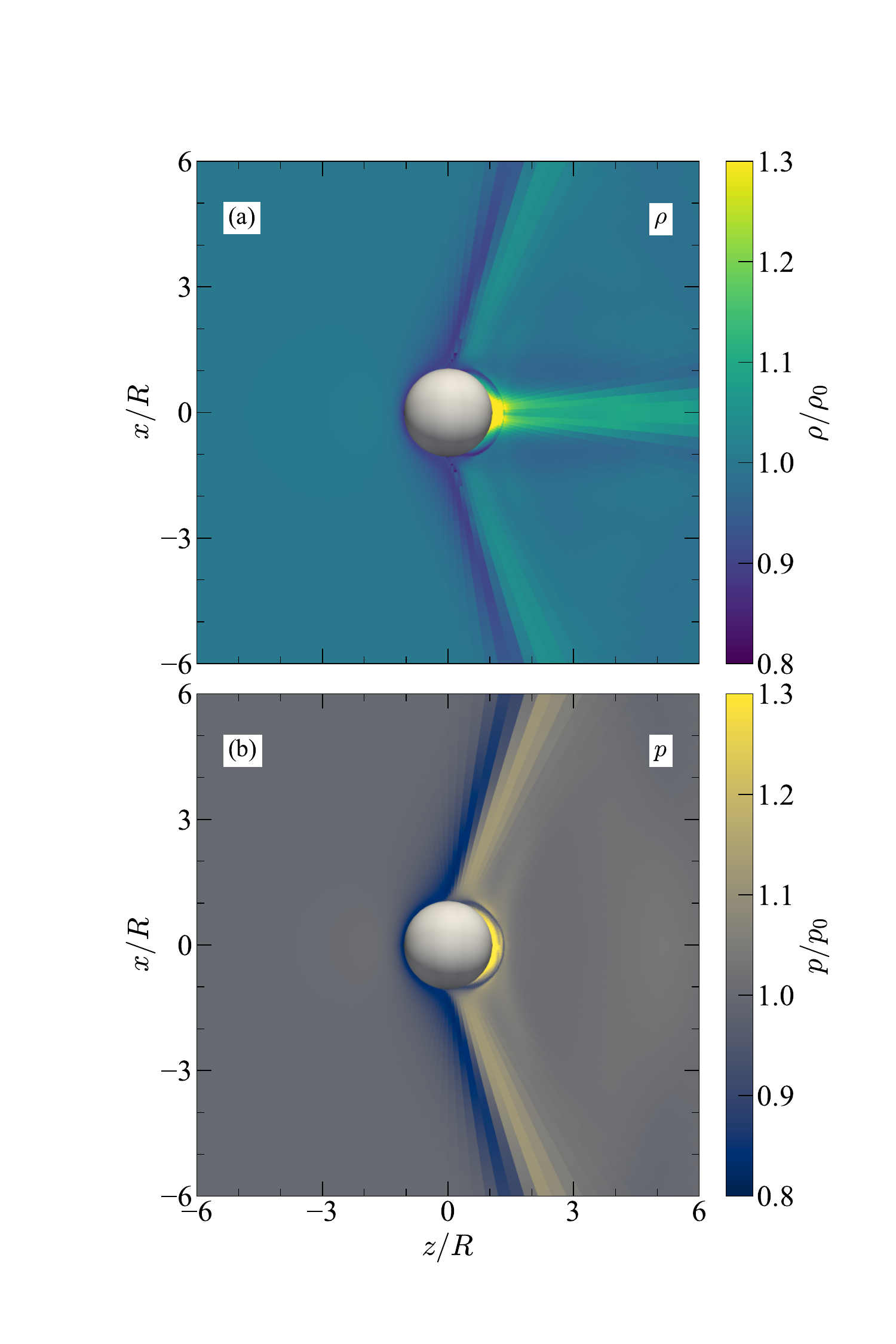}
    \caption{Density (a) and pressure (b) for our fiducial run, with
    $\MA=0.3$, $\beta=2$, on the \plane{x}{z} after the
    simulation has achieved a steady state. \label{fid_visual_2}}
  \end{center}
\end{figure}

\subsection{Analysis and Observables}

We are interested in measuring the drag force on the conductor. A direct
evaluation of this quantity may be obtained by taking a volume integral of the
rate of change of momentum density (see Eq.
\ref{momentum}),
\begin{multline}
  \vb{F_\mathrm{drag}} = \int\pdv{\vb{m}}{t}\dd V \\
    = - \oint \left(\vb{m} \vb{v}-\frac{\vb{\dB} \vb{B}}{4\pi}
        -\frac{\vb{\Bbkg} \vb{\dB}}{4\pi}+ p_t \vb{\mathbf{I}}\right)\vdot\dd\vb{A}. \label{F_drag}
\end{multline}
To evaluate Eq. (\ref{F_drag}), we choose our surface to be a sphere that
encloses the conductor. We evaluate this drag force in terms
of the mechanical power imparted onto the conductor,
\textit{i.e.}, $\dot{E}=\vb{F_\mathrm{drag}}\vdot\vb{v_0}=F_\mathrm{drag,z}v_0$.
Note that we adopt the convention that $\dot{E}>0$ corresponds to a drag that
slows down the object in the plasma rest frame. Evaluating Eq. (\ref{F_drag})
at radius $r$, we find
\begin{multline}
  \dot{E}(r)=-r^2v_0\oint\left(m_z v_r-\frac{\delta B_z B_r}{4\pi} \right. \\
               \left. -\frac{B_{\mathrm{bkg},z} \delta B_r}{4\pi} + p_t \cos\theta\vphantom{\frac{\delta B_z B_r}{4\pi}}\right)\dd\Omega. \label{energy_flux}
\end{multline}

The \Alfven wing model \citep{Drell+65} suggests that the power imparted onto
the conductor should derive from the perturbed Poynting flux propagating along
the \Alfven wings. For $v_0\lesssim \vAn$, the \Alfven wave amplitude at $r\sim R$ is of
order $\flatfrac{B_0 v_0}{\vA}$, and the wing has an area $\sim\pi R^2$. Thus
the power is
\begin{multline}
\dot{E}\sim\frac{1}{4\pi}\left(B_0\frac{v_0}{\vAn}\right)^2\vAn \left(2\pi R^2\right) \\
= \frac{1}{2} B_0^2 R^2 \frac{v_0^2}{\vAn} = 2\rho_0\pi R^2 v_0^2 \vAn. \label{Edot}
\end{multline}
Motivated by Eq. (\ref{Edot}), we define a dimensionless drag coefficient
$\CA$, such that
\begin{equation}
\dot{E}=\CA \rho_0 R^2 v_0^2 \vAn, \label{drag_coef}
\end{equation}
with $\CA=\CA(\MA,\beta)$ depending only on $\MA$ and $\beta$ (see Eq.
\ref{params}).

In our study, we run a given simulation until $\CA$ reaches a constant value,
indicating that the simulation has reached a steady state. This usually takes
several crossing times, $t_\mathrm{cross}=\flatfrac{R}{v_0}$, to achieve but
may occur more rapidly.  For this study, we run simulations for
$\MA\in[0.003,1]$ spaced logarithmically, and for $\beta=200, 2,\text{and }
0.02$. We also perform additional runs for $\MA=0.01,0.1$ and
$\beta\in[0.002,200]$ spaced logarithmically to focus on the scaling of $\CA$
with $\beta$ in the linear regime. Since some of these cases approach
supersonic speeds, we switch to the HLLD Riemann solver when $\MA\geq0.1$ for
$\beta=0.02$ and when $\MA\geq0.7$ for $\beta=2$ to improve the robustness of
the simulations. Note that when $\beta=0.02$ and $\MA\geq0.7$, the simulations
fail due to the development of a strong evacuated region in front of the
conductor, leading to failures in our numerical method.

\section{Results} \label{sec:results}

\subsection{Fiducial Run}

\begin{figure}[ht] 
  \begin{center}
    \includegraphics[width=\columnwidth, trim={100 50 25 100}, clip]{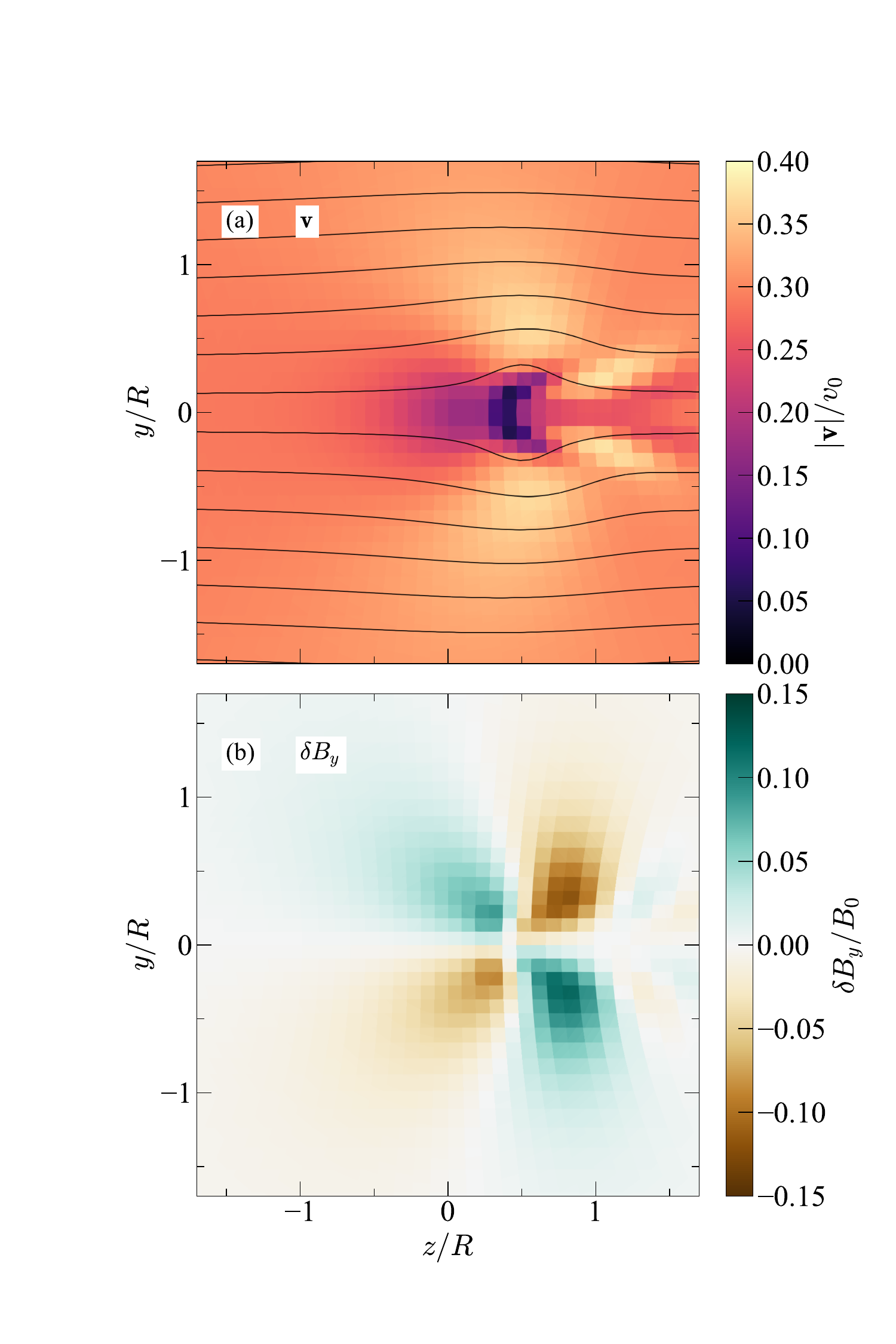}
    \caption{Diagnostic parameters for our fiducial run on the $x=-2R$ plane
    (seen from below) after the simulation has achieved a steady state. Shown
    are (a) $\vb{v}$ and (b) $\delta B_y$. The vector field lines for $\vb{v}$
    initialize at $x=2.5R$ since there is a strong out-of-plane component
    close to the wing [see Figure \ref{fid_visual}(c)]. \label{fid_yz_visual}}
  \end{center}
\end{figure}

Our fiducial run has $\MA=\flatfrac{v_0}{\vAn}=0.3$ and $\beta=2$. In Figs.
\ref{fid_visual} and \ref{fid_visual_2}, we depict several diagnostic variables
after a steady state has been achieved. The evolved quantities ($\delta\vb{B}$
and $\vb{v}$) show noticeable perturbations along the \Alfven wing
characteristics, which are defined as
\begin{equation}
  \vb{c_\mathrm{A}^\pm} = \vb{v} \pm \vb{\vA}, \label{characteristics}
\end{equation}
where $\vb{\vA}\equiv\flatfrac{\vb{B}}{\sqrt{4\pi\rho}}$. On the front side of
the wing, we observe an enhancement of the magnetic field and a reduction in
the gas density and pressure. Directly along $\vb{c_\mathrm{A}^\pm}$, we
observe a narrow region that has nearly zero velocity. This region co-moves
with the conductor, and the enhanced magnetic flux is frozen in, creating a
narrow obstacle that the background plasma passes around. This behavior is
highlighted in Figure \ref{fid_yz_visual}, which shows some of the diagnostics
in the $x=-2R$ plane. The flow drags the magnetic field lines to form loops
around the \Alfven wing.

Figures \ref{fid_visual}--\ref{fid_yz_visual} also reveal some features not
directly associated with the \Alfven wings. For instance, immediately
downstream of the conductor is a small bubble which exhibits enhanced magnetic
field strength, density, and pressure. This region is partially trapped,
confined by magnetic field lines wrapping around the conductor. We do observe a
density enhancement downstream of this feature, but it is transported by the
background flow, so in the cases of low $\MA$, it carries significantly less energy
away from the conductor compared to the \Alfven wings.

\begin{figure}[ht]
  \begin{center}
    \includegraphics[width=\columnwidth, trim={50 0 75 25}, clip]{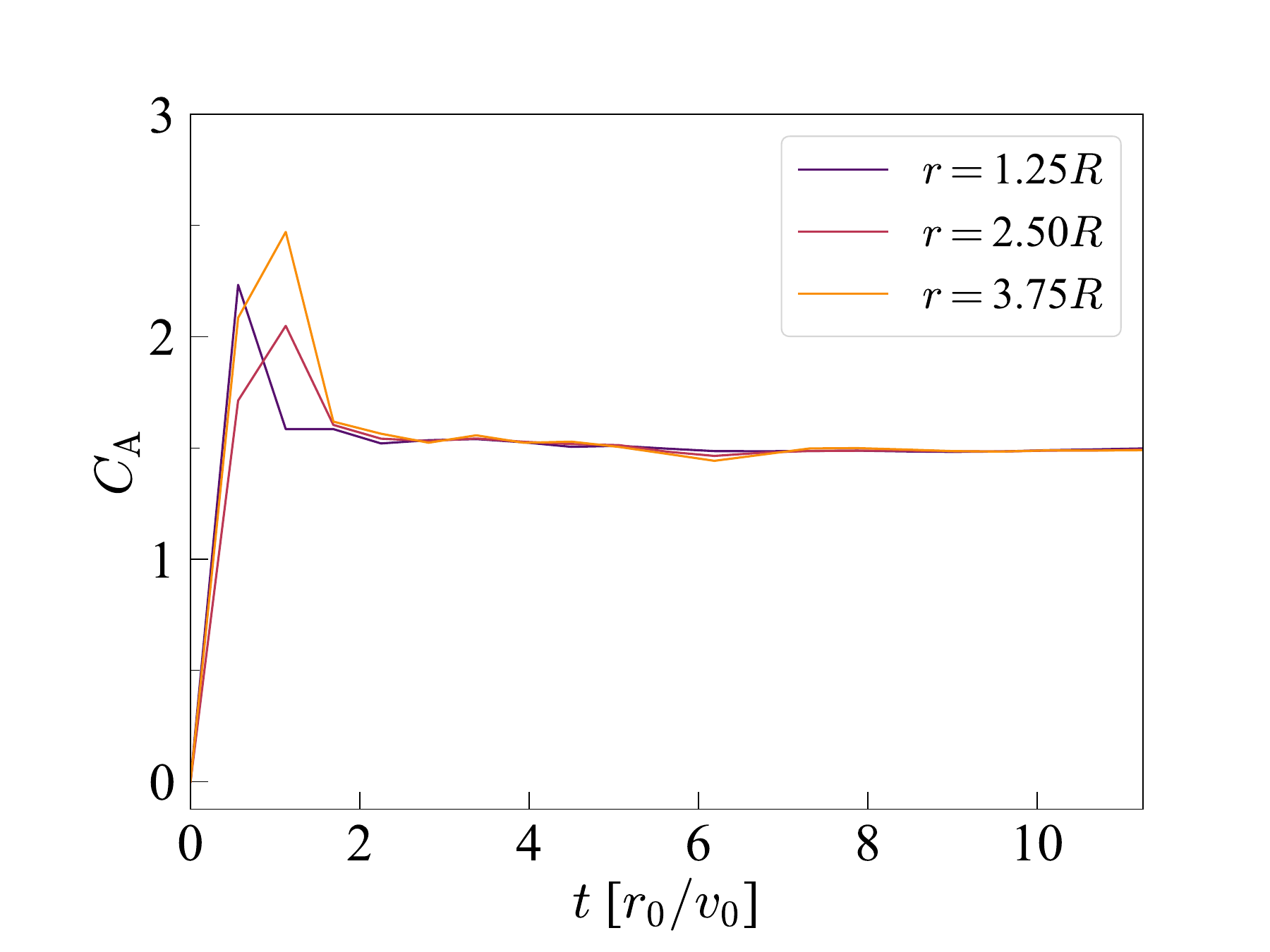}
    \caption{Time evolution of the drag coefficient $\CA$ (see Eq. \ref{drag_coef}) measured at different radii (see Eq. \ref{energy_flux})
    in our fiducial run ($\MA=0.3$ and $\beta=2$). For all simulations, the system is evolved until the
    value for $\CA$ stabilizes, as shown in this case. $\CA$ is evaluated at
    $r=\{1.25, 2.50, 3.75\} R$ (purple, magenta, and orange, respectively). \label{fid_ev}}
  \end{center}
\end{figure}

Figure \ref{fid_ev} depicts the drag coefficient $\CA$ (see Eq.
\ref{drag_coef}) as a function of time for our fiducial run. After an initial
transient phase, we find that the simulation rapidly arrives at a steady state,
where evaluations of $\CA$ at different radii achieve the same value. We use
this level of agreement as a diagnostic for the quality of the simulation, and
runs that do not converge to a single value of $\CA$ over time are rerun at
higher resolution.

\begin{figure}[ht]
  \begin{center}
    \includegraphics[width=\columnwidth, trim={50 0 75 50}, clip]{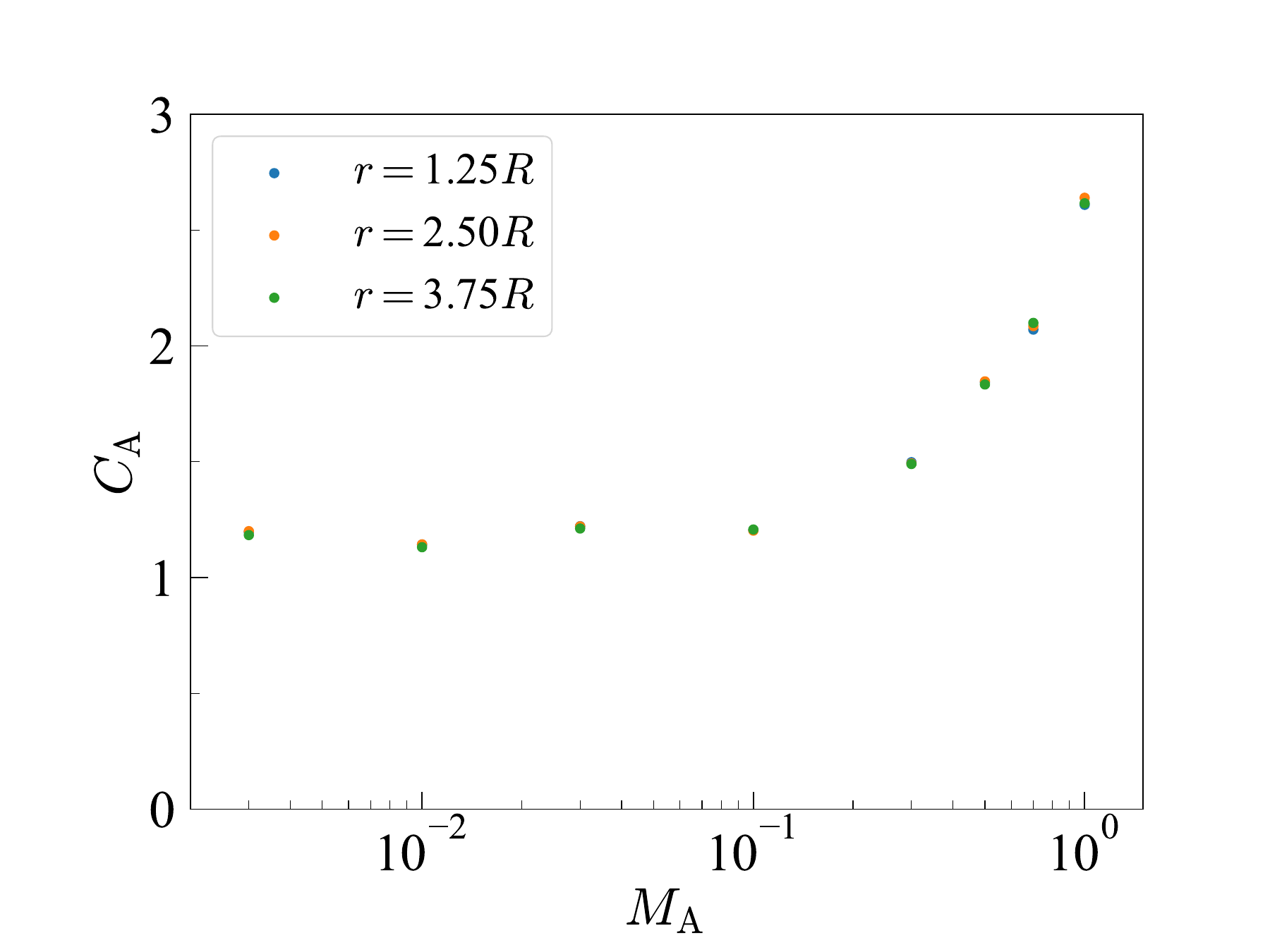}
    \caption{The drag coefficient $\CA$ measured at different radii as a function of
    $\MA$ for $\beta=2$. The
    values are obtained following the procedure outlined in Figure
    \ref{fid_ev}. The residual spread among different radii is interpreted as a
    level of ``error'' in our measurement. The value of $\CA$ is roughly
    constant for sub-\Alfvenic speeds, but noticeably increases as
    $\MA$ approaches unity. \label{v0_comp}}
  \end{center}
\end{figure}

\begin{figure}[ht]
  \begin{center}
    \includegraphics[width=\columnwidth, trim={50 0 75 25}]{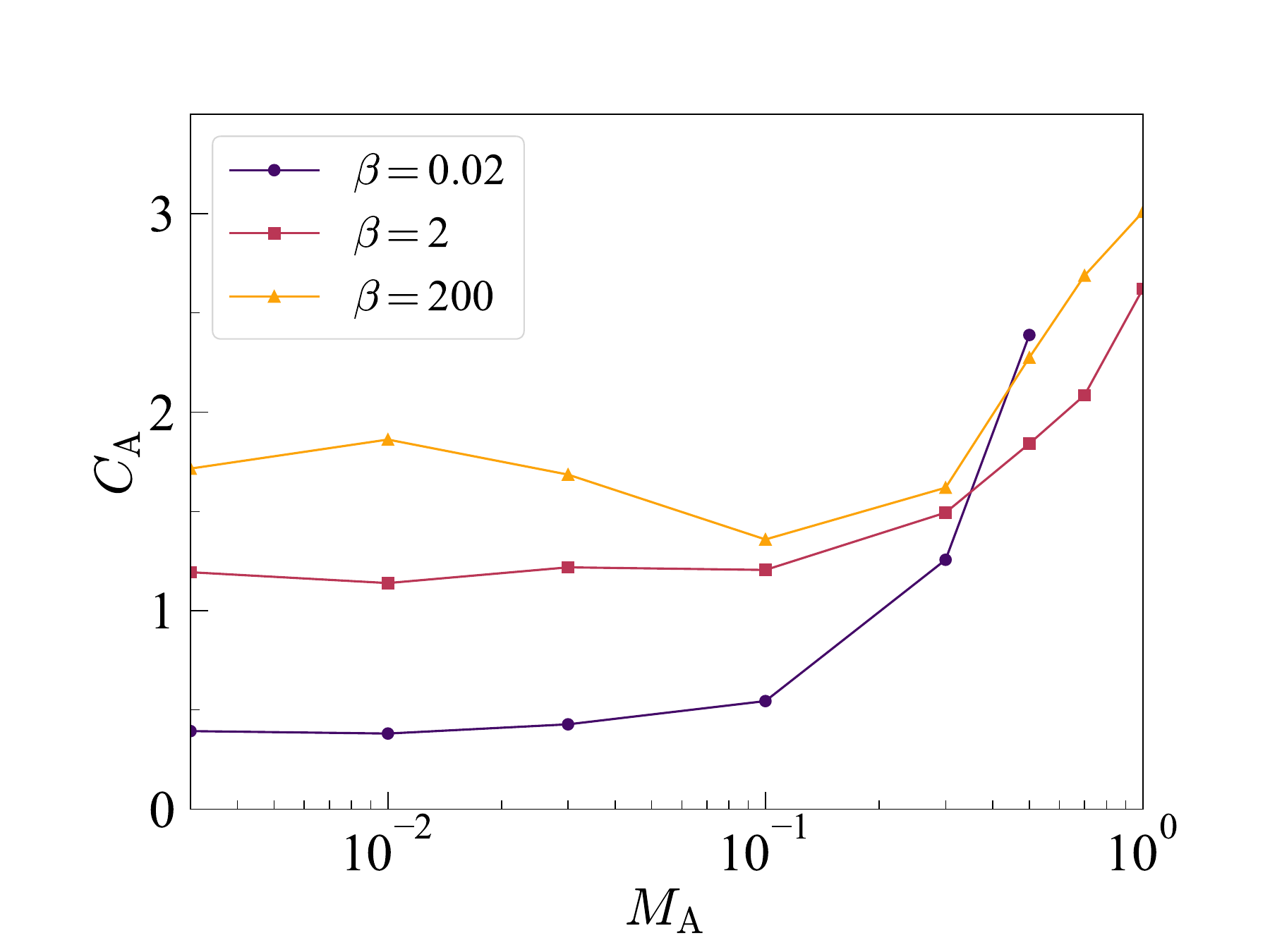}
    \caption{The drag coefficient $\CA$ measured at $r=1.25R$ as a function of
    $\MA$ for $\beta=0.02,2,200$ (purple circle, magenta square, yellow triangle).
    \label{beta_comp}}
  \end{center}
\end{figure}

\begin{figure}[ht]
  \begin{center}
    \includegraphics[width=\columnwidth, trim={25 0 75 50}, clip]{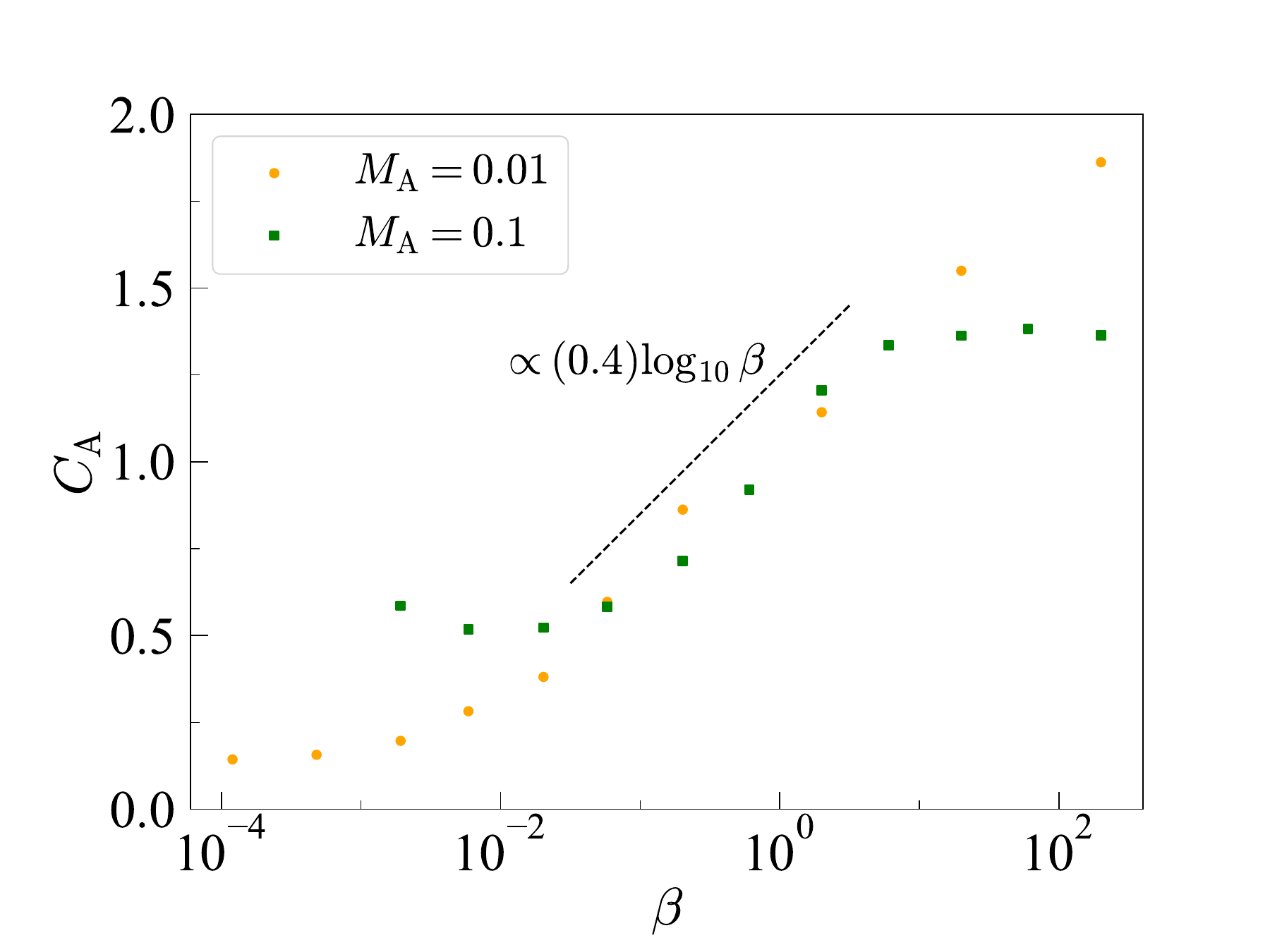}
    \caption{The drag coefficient $\CA$ as a function of $\beta$, for
    $\MA=0.01$ (yellow circle) and $\MA=0.1$ (green square). The black dashed
    line indicates the scaling that occurs in the regime where $\CA$ is
    increasing. \label{v0.01_beta_comp}}
  \end{center}
\end{figure}

\begin{figure*}[!htb]
  \begin{center}
    \includegraphics[width=\textwidth, trim={100 125 0 150}, clip]{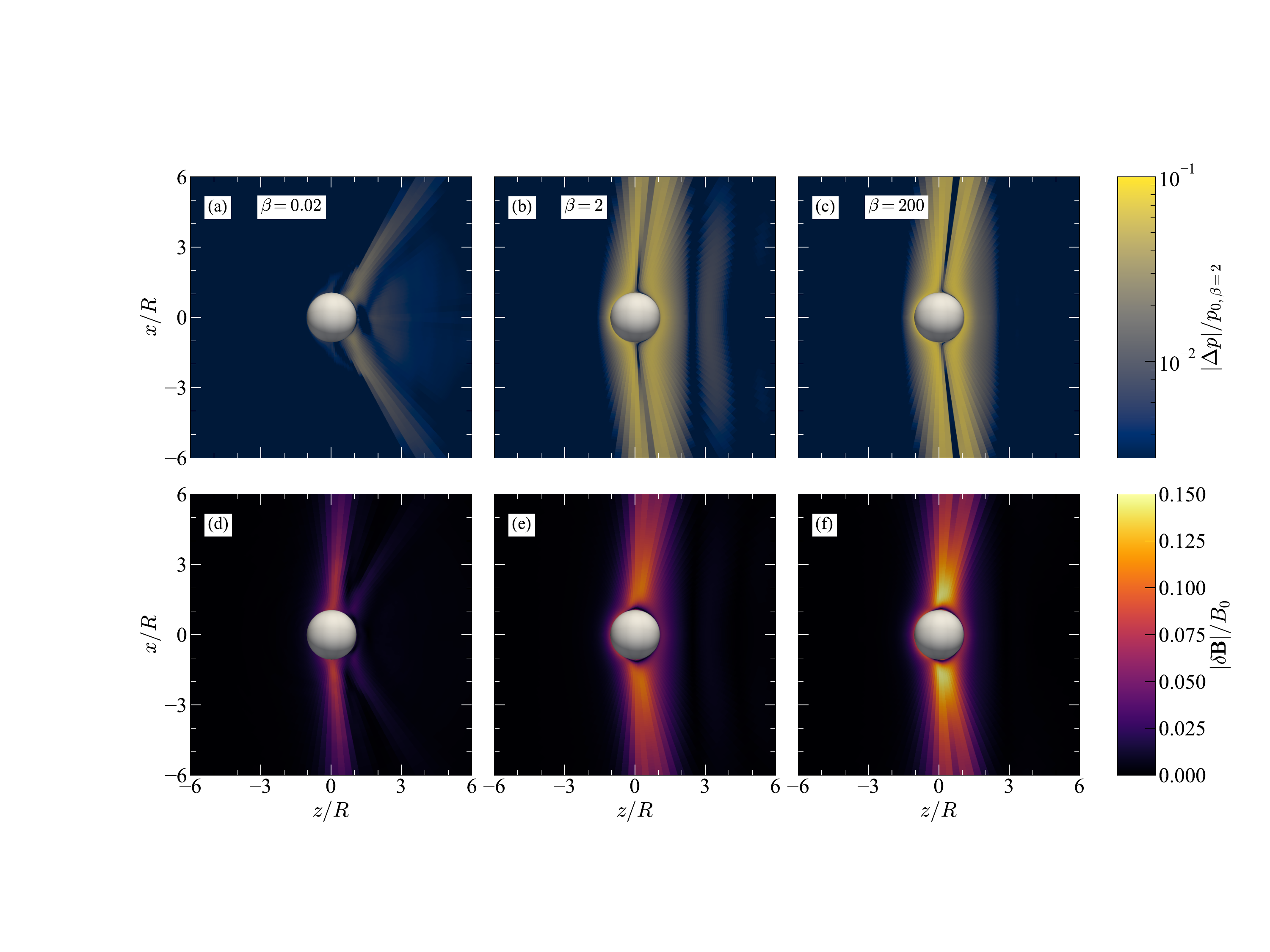}
    \caption{Gas pressure and magnetic field disturbances for $\MA=0.1$, with different
    values of $\beta$ on the \plane{x}{z} after each simulation has achieved
    a steady state. The top row shows $|\Delta p|=|p-p_0|$, normalized by the
    background pressure of the $\beta=2$ case, and the bottom row shows $|\vb{\delta B}|$.
    \label{beta_comp_visual}}
  \end{center}
\end{figure*}

\begin{figure*}[!htb]
  \begin{center}
    \includegraphics[width=\textwidth, trim={75 25 50 25}, clip]{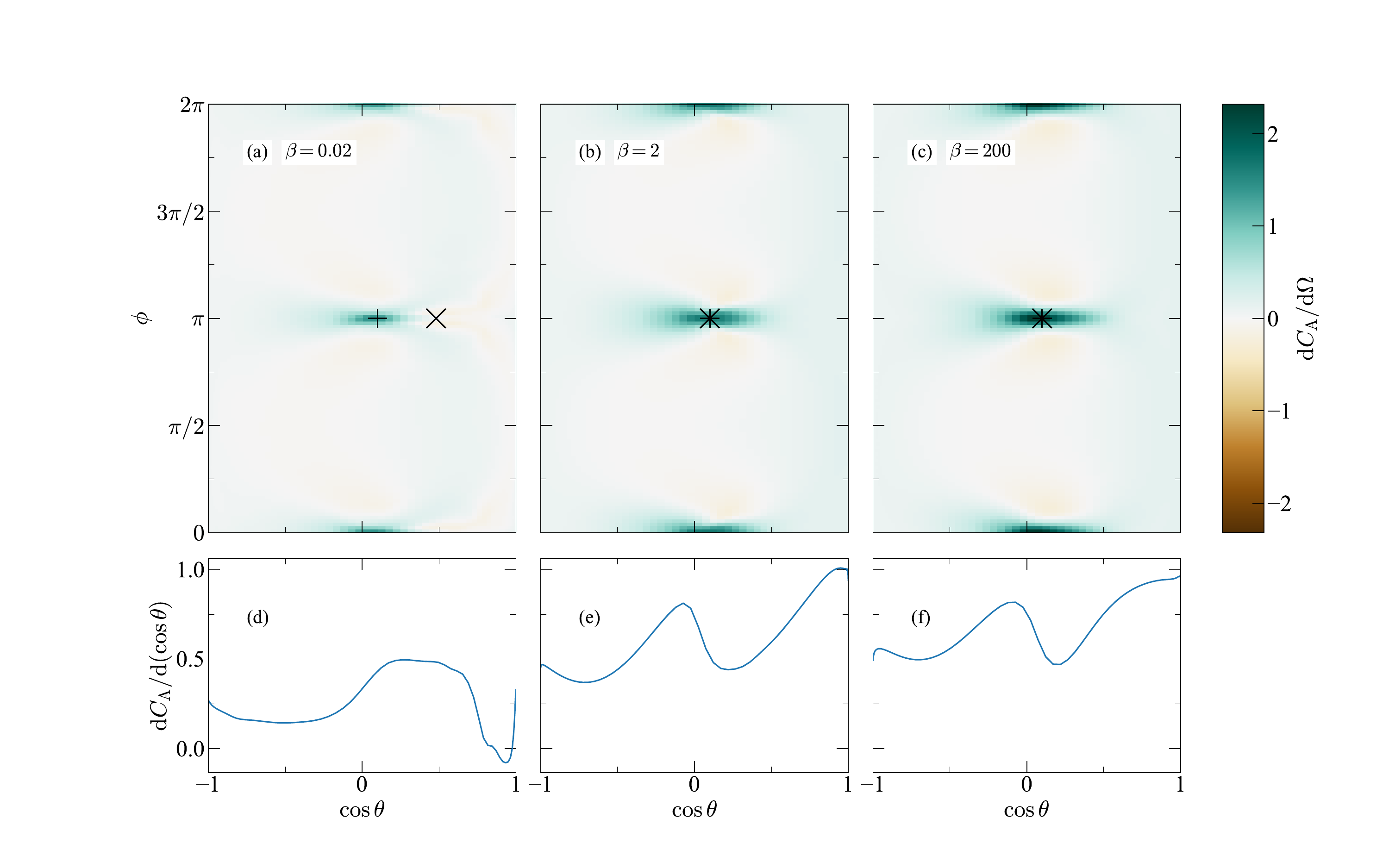}
    \caption{Distribution of the contributions to $\CA$ measured at $r=1.5R$ over
    solid angle (top row) and only $\cos\theta$ (bottom row) for $\MA=0.1$ with
    different values of $\beta$. The contributions from the background
    flow, defined in Eq. (\ref{background}), were subtracted out to
    highlight new features in the steady state. The black plus signs and
    crosses indicate the locations of the \Alfven and slow wing
    characteristics, respectively. \label{maps_v0.1}}
  \end{center}
\end{figure*}

\begin{figure*}[!htb]
  \begin{center}
    \includegraphics[width=\textwidth, trim={75 25 50 25}, clip]{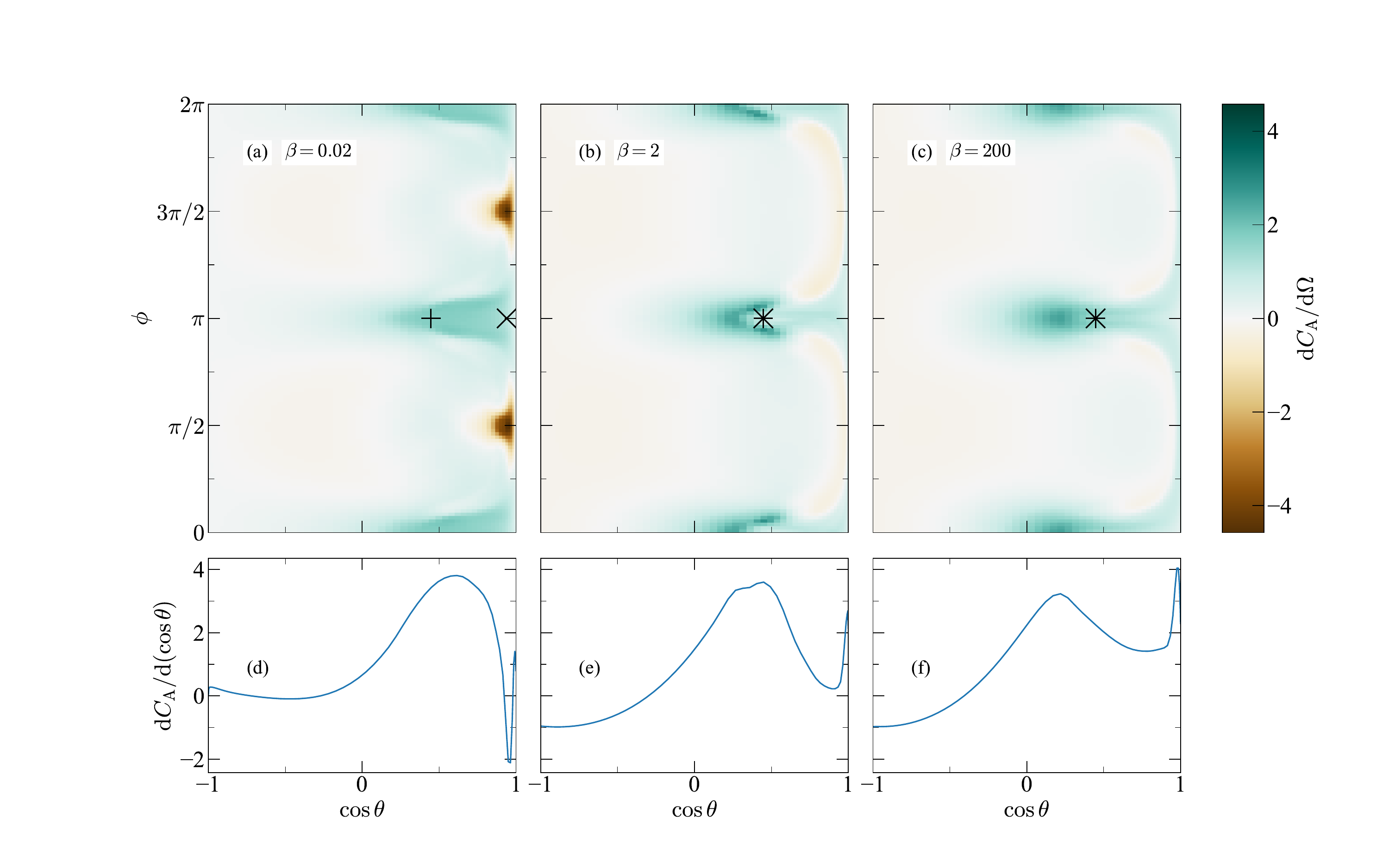}
    \caption{Distribution of contributions to $\CA$ measured at $r=1.5R$ over
    solid angle (top row) and only $\cos\theta$ (bottom row) for $\MA=0.5$ with
    different values of $\beta$. The contributions from the background
    flow, defined in Eq. (\ref{background}), were subtracted out to
    highlight new features in the steady state. The black plus signs and
    crosses indicate the locations of the \Alfven and slow wing
    characteristics, respectively. \label{maps_v0.5}}
  \end{center}
\end{figure*}

\subsection{$\MA$- and $\beta$-Dependence}

Figure \ref{v0_comp} depicts $\CA$ as a function of $\MA$ at our fiducial
$\beta=2$. For low values of $\MA$, the drag coefficient remains
roughly constant, at a value $\CA\approx1.2$, indicating a strong agreement
with the \Alfven wing model in this regime. However, for $\MA\gtrsim 0.1$, we
find a noticeable increase in $\CA$. In this case, the system enters a
nonlinear regime, where the wings point so far forward that they disturb the
flow downstream of the conductor.

Figure \ref{beta_comp} shows the $\CA$ vs. $\MA$ relations for different values
of $\beta$. For each case, we see the same general pattern, where sub-\Alfvenic
flows exhibit a constant value of $\CA$, in good agreement with the standard
\Alfven wing picture, but flows with larger $\MA$ exhibit a noticeable increase
in the drag coefficient. For the $\beta=0.02$ case, we find that this
transition to a higher drag sets in at lower values of $\MA$. Indeed, the
upward swing of $\CA$ occurs when $v_0\approx \csn$. Additionally, even in the
sub-\Alfvenic regime where the \Alfven wing model appears robust, we find that
$\CA$ strictly increases with $\beta$.

Figure \ref{v0.01_beta_comp} focuses on this sub-\Alfvenic regime by plotting
$\CA$ as a function of $\beta$ for $\MA = 0.01$ and $\MA=0.1$. For low values
of $\beta$, there is no strong relationship between $\CA$ and $\beta$, but
after a transition point, we find that the $\CA$--$\beta$ relation is well
described by $\CA\simeq(0.4)\log_{10}\beta$. The transition point where this
scaling begins varies between the two cases but corresponds to the point where
the condition $\MA\simeq\flatfrac{\cs}{\vA}$ holds. We note that the $\MA=0.1$
case shows a high $\beta$ regime where this trend again flattens off, but that
this is absent in the $\MA=0.01$ case.

To explore this trend with $\beta$ further, we show the disturbances in both
the gas pressure and the magnetic field in Figure \ref{beta_comp_visual} for
several runs with different values of $\beta$, all with $\MA=0.1$. We see that
for the $\beta=0.02$ case, the magnetic disturbance and pressure disturbance
have very different patterns. While the former still follows the \Alfven
characteristic (Eq. \ref{characteristics}), the latter follows a more
``inclined'' trajectory, corresponding approximately to the characteristic
\begin{equation}
  \vb{c_{-}^\pm} = \vb{v} \pm \cs\vu{x},
\end{equation}
where $\cs=\left(\flatfrac{\gamma p}{\rho}\right)^{1/2}$ is the gas sound
speed. The ``high inclination'' of the pressure disturbance for the low $\beta$
case suggests that it is generated by the slow magnetosonic wave, which is
described by the dispersion relation,
\begin{equation}
\left(\frac{\omega}{k}\right)^2 = \frac{1}{2} \left(\vA^2 +
\cs^2 - \sqrt{\left(\vA^2 +
\cs^2\right)^2-4\vA^2\cs^2\cos^2\alpha}\right),
\end{equation}
where $\alpha$ is the angle between the wavevector $\vb{k}$ and the magnetic
field, and $\omega$ is the angular frequency.
Unlike \Alfven waves, which propagate at a group velocity aligned with the
magnetic field, the slow waves have some degree of dispersion, so they have
been neglected in the context of magnetic interactions in binary systems since
they are believed to be unfocused and unrelated to the \Alfven wing picture
\citep{Strugarek23, Neubauer80}. However, in the cases of extreme $\beta$, the slow
wave group velocity may still effectively collimate in a single direction,
approaching $\vb{v_\mathrm{slow}}\approx \cs\vu{b}$ in the case of $\beta\ll1$
and $\vb{v_\mathrm{slow}}\approx \vA\vu{b}$ in the case of $\beta\gg1$, where
$\vu{b}$ is the unit vector aligned with the magnetic field. This suggests that
the slow wave could also produce a wing-like structure which is distinct from
the \Alfven wing in the former case but collinear in the latter case, which is
precisely what occurs in Figure \ref{beta_comp_visual}.

Figure \ref{maps_v0.1} shows the angular distribution of the contributions to
$\CA$ for the same three cases depicted in Figure \ref{beta_comp_visual}:
\begin{equation}
\pdv{\CA}{\Omega}=\frac{1}{\rho_0 R^2 v_0^2 \vAn} \pdv{E}{\Omega},
\end{equation}
where we have, from Eq. (\ref{energy_flux}),
\pagebreak
\begin{multline}
\pdv{\CA}{\Omega}=-r^2 v_0\left(m_z v_r-\frac{\delta B_z B_r}{4\pi}  \right. \\
\left. -\frac{B_{\mathrm{bkg},z} \delta B_r}{4\pi} + \vphantom{\frac{\delta B_z B_r}{4\pi}} p_t \cos\theta\right).
\end{multline}
To highlight interesting features, the contributions from the background flow
are subtracted; \textit{i.e.}, since we have $\eval{\delta B}_{t=0}=0$, we may
apply Equation (\ref{energy_flux}) to find the initial drag contribution,
\begin{equation}
\eval{\pdv{\dot{E}(r)}{\Omega}}_{t=0}=-r^2v_0\left(\rho_0 v_0^2 + p_0
\cos\theta \right). \label{background}
\end{equation}
Due to the symmetry of the initial data, Equation (\ref{background}) contains a
dipolar contribution, which cancels itself out when integrating over the
sphere. As such, we lose no important information by subtracting this component
out. For what remains we find that much of the energy dissipation is indeed
being propagated through the \Alfven wings, but in the case of higher plasma
$\beta$ we see strong contributions at the polar regions as well. This is
likely due to the pressure enhancements building up in front of and behind the
sphere. We note that for the $\beta=0.02$ case, the direction of the slow wave
points to a collimated region which contributes little to the drag force, and
may even hinder it.

Figure \ref{maps_v0.5} shows the same quantities for the case of $\MA = 0.5$.
In this case, the power dissipated by the disturbance is less collimated along
the \Alfven characteristics, instead filling the entire region downstream of
the conductor. This reflects the nonlinear structure of the wings in the
immediate vicinity of the sphere, which modifies the effective size of the
obstacle. At larger distances the disturbances would again collimate along the
\Alfven characteristics, but the size of the wings reflects the new effective
size of the obstacle. This larger disturbance enhances the overall drag imposed
on the sphere, resulting in the enhancement seen in Figure \ref{v0_comp}.

\section{Summary and Conclusions} \label{sec:discuss}

We have performed 3D MHD simulations of magnetized flow past an
unmagnetized spherical conductor. By numerically calculating the momentum flux
carried away by the flow and magnetic disturbances along the \Alfven wings, we
determine the drag force on the conductor. The drag coefficient $\CA$ is a
function of two dimensionless parameters, the \Alfven Mach number $\MA=\flatfrac{v_0}{\vAn}$, and
the plasma $\beta=\flatfrac{\left(\flatfrac{\csn^2}{\gamma}\right)}{\vAn^2}$. Our
simulations reveal distinct \Alfven wing features, and we find consistency with
the \Alfven wing model at sub-\Alfvenic speeds. Significant deviations occur
when $\MA$ approaches unity (see Figs. \ref{v0_comp}--\ref{beta_comp}). We also
find a positive correlation between $\beta$ and $\CA$ (see Fig.
\ref{v0.01_beta_comp}).

The latter result is likely the consequence of a secondary wing associated with
the slow magnetosonic wave, which is highly collimated for plasma $\beta$ far
from order unity. When $\beta \ll 1$, the slow magnetosonic wave essentially
travels at the speed of sound along magnetic field lines, and when $\beta \gg
1$, it travels at the \Alfven speed. As such, in the low $\beta$ environments,
this `slow' wing and its corresponding pressure disturbance separates from the \Alfven wing. Meanwhile, the magnetic disturbance remains concentrated along the \Alfven wing, so it is unclear
what sorts of unique observational consequences the slow wing would have. For high
$\beta$ environments, the two wing features are collinear, producing a single enhanced disturbance along the \Alfven wing characteristic.

There are several direct extensions of this work which we intend to pursue.
Most compelling are the effects of the internal magnetic field of the
conductor, $\vb{\Bin}$. It may give rise to additional magnetic dissipation in
the near vicinity of the conductor. There have already been some investigations
into this in the context of star-planet (\textit{e.g.} \citealt{Strugarek+14,
Strugarek+15}) and compact object binary (\textit{e.g.}
\citealt{Most&Philippov22, Most&Philippov23}) systems. How the
drag force depends on $\vb{B_0}$, $\vb{\Bin}$, and other parameters remains
unclear.

We also note that our study is currently confined to ideal MHD, so we do not
consider the effects of viscosity or resistivity. Viscosity would introduce a
boundary layer around the sphere. Resistivity would enable magnetic
reconnection, which could dissipate energy from the \Alfven wings at their
boundaries with the surrounding medium. \cite{DeColle+25} include resistivity
in their magnetized flow simulations, but it remains an open question how
varying the parameter affects the system. Consideration of these features would
introduce the Reynolds number and the magnetic Reynolds number as dimensionless
canonical parameters of this problem. Indeed, this would bring the system
closer to the original context that inspired this work, viscous flow past a
sphere. 

\begin{acknowledgments} This research was partially supported by the
NSF grant DGE-2139899. \software{\texttt{Matplotlib}
\citep{Hunter07}, \texttt{NumPy} \citep{Harris+20}, \texttt{ParaView}
\citep{Ayachit15}.} \end{acknowledgments}


\bibliography{Bibliography}{} \bibliographystyle{aasjournal}

\end{document}